\documentclass[11pt]{article}

\usepackage[utf8]{inputenc}
\usepackage{amsmath}
\usepackage{amssymb}
\usepackage{amsthm}
\usepackage{color,soul}
\usepackage{tikz}
\usetikzlibrary{topaths,calc}
\usepackage{subfigure}
\usepackage{cite}

\usepackage{arydshln} 
\usepackage{booktabs}  
\usepackage{amssymb}
\usepackage{amsmath}

\def\QED{\mbox{\rule[0pt]{1.5ex}{1.5ex}}}

\def\QED{\mbox{\rule[0pt]{1.5ex}{1.5ex}}}

\newtheorem{theorem}{Theorem}

\newtheorem{lemma}{Lemma}

\newtheorem{remark}{Remark}

\newcommand\blfootnote[1]{%
  \begingroup
  \renewcommand\thefootnote{}\footnote{#1}%
  \addtocounter{footnote}{-1}%
  \endgroup
}

\title{Vector Linear Secure Aggregation}
\author{Xihang Yuan and Hua Sun}
\date{}

\begin{document}
\maketitle

\blfootnote{
Xihang Yuan (email: xihangyuan@my.unt.edu) and Hua Sun (email: hua.sun@unt.edu) are with the Department of Electrical Engineering at the University of North Texas.}

\begin{abstract}
The secure summation problem, where $K$ users wish to compute the sum of their inputs at a server while revealing nothing about all $K$ inputs beyond the desired sum, is generalized in two aspects - first, the desired function is an arbitrary linear function (multiple linear combinations) of the $K$ inputs instead of just the sum; second, rather than protecting all $K$ inputs, we wish to guarantee that no information is leaked about an arbitrary linear function of the $K$ inputs. For this vector linear generalization of the secure summation problem, we characterize the optimal randomness cost, i.e., to compute one instance of the desired vector linear function, the minimum number of the random key variables held by the users is equal to the dimension of the vector space that is in the span of the vectors formed by the coefficients of the linear function to protect but not in the span of the vectors formed by the coefficients of the linear function to compute.
\end{abstract}

\newpage
\allowdisplaybreaks
\section{Introduction}
Secure summation, motivated in particular by federated learning \cite{aggregation, Zhao_Sun_Aggregate, aggregation_light}, studies the information theoretic limits on computing the sum of inputs from distributed users without revealing any additional information to the aggregating server. Specifically, suppose we have $K$ users where User $k \in \{1,2,\cdots,K\}$ holds an input $W_k$ and an independent key $Z_k$. 
To securely compute the sum $W_1 + \cdots + W_K$, each user sends a message $X_k$ to the server so that from $X_1, \cdots, X_K$, the server may recover $W_1+ \cdots + W_K$ while learns no additional information about all inputs $W_1, \cdots, W_K$ beyond the sum. In \cite{Zhao_Sun_Sum}, the fundamental limit on communication and randomness cost of secure summation is characterized - to securely compute $1$ bit of the sum, each user needs to send $1$ bit of message $X_k$ to the server, each user needs to hold $1$ bit of key $Z_k$ and all key variables $Z_1, \cdots, Z_K$ need to have $K-1$ bits of entropy (randomness).

In this work, we consider a vector linear generalization of the above secure summation problem where both the desired function to compute and the security function to protect are arbitrary linear transforms - first, instead of computing the sum, the users may aggregate the inputs through an arbitrary linear function ${\bf F}[W_1; \cdots; W_K]$ where ${\bf F}$ is a $\mbox{rank}({\bf F}) \times K$ matrix and $[{\bf A}; {\bf B}]$ is the row stack of matrices ${\bf A}, {\bf B}$; second, rather than protecting all inputs, the users may guarantee nothing is revealed about an arbitrary linear function of the inputs, ${\bf G}[W_1; \cdots; W_K]$ where ${\bf G}$ is a $\mbox{rank}({\bf G}) \times K$ matrix. 
The relaxation of protecting all source information to a certain (linear) function of the sources has also been recently considered in the context of network coding for computing \cite{Xuan_Yang_Yeung_Source, Xuan_Yang_Yeung_Function}.

The main result of this work is the characterization of the information theoretic minimum randomness cost for vector linear secure aggregation. 
We show that the key variables $Z_1, \cdots, Z_K$ need to have entropy of $\mbox{rank}([{\bf F}; {\bf G}]) - \mbox{rank}({\bf F}) \triangleq \mbox{rank}({\bf G}|{\bf F})$ bits per input bit. 
The result manifests itself in interpretation - we need key variables to cover the row space of ${\bf G}$ to avoid information leakage while the row space of ${\bf F}$ does not need to be protected as it must be correctly computed anyway so the keys are only required to cover the space that lies in ${\bf G}$ but not in ${\bf F}$, i.e., the answer can be expressed as the conditional rank term. 
To prove this result, the key challenge is on the achievability side while the converse side follows in a relatively straightforward manner from \cite{Zhao_Sun_Sum}. 
In particular, the crux is how to design the key variables so that we are only protecting the part of space that is in ${\bf G}$ while out of ${\bf F}$ and at the same time ensure the desired function can be recovered with no error. 
Interestingly, our scheme is inspired by entropic analysis, i.e., we are guided by the entropy condition of the security constraint (refer to (\ref{security})) and reverse engineer to construct the code so that the mutual information in the security constraint is forced to be zero (refer to Section \ref{sec:ach}). 
The total randomness rate result of secure summation is recovered as a special case, as here ${\bf G}$ is the $K \times K$ identity matrix and ${\bf F}$ is the all $1$ row vector so that $\mbox{rank}([{\bf F}; {\bf G}]) - \mbox{rank}({\bf F}) = K-1$. 
Lastly, a few comments on other performance metrics. The communication cost of vector linear secure aggregation is the same as that of secure summation, i.e., each user still needs to send all information contained in $W_k$ through $X_k$ to the server. 
The individual key rate (i.e., the minimum size of $Z_k$) turns out to be an open problem. Note that for secure summation, each individual input $W_k$ needs to be fully protected so that the size of $Z_k$ cannot be less than that of $W_k$ while this stringent security constraint is relaxed in vector linear secure aggregation and we are not required to protect each individual $W_k$ but we are not able to determine how much leakage is allowed exactly in general. 
For a more detailed discussion, refer to Section \ref{sec:dis}.


\section{Problem Statement}\label{sec:model}
Consider $K$ users and User $k$ $\in$ $\{1,2,\cdots,K\} \triangleq [K]$ holds input $W_k$ and key $Z_k$. 
The inputs $(W_k)_{k\in[K]}$ are independent and are independent of $\left(Z_k\right)_{k\in[K]}$.
Each $W_k$ is a $1 \times L$ row vector with i.i.d. uniform elements from the finite field $\mathbb{F}_q$. 
\begin{eqnarray}
    && H\left(\left( W_k \right)_{k\in [K]},\left( Z_k \right)_{k\in[K]}\right)=\sum_{k\in[K]}H\left( W_k \right)+H\left(\left( Z_k \right)_{k\in[K]}\right), \label{h1}\\
    && H\left( W_k \right) = L ~(\mbox{in $q$-ary units}), ~\forall k \in [K]. \label{h2}
\end{eqnarray}
Each $Z_k$ is comprised of $L_Z$ symbols from $\mathbb{F}_q$. $\left( Z_k \right)_{k\in[K]}$ are a function of a source key variable $Z_{\Sigma}$, which is comprised of $L_{Z_{\Sigma}}$ symbols from $\mathbb{F}_q$.
\begin{eqnarray}
    H\left(\left(Z_k\right)_{k\in[K]}\Bigl|Z_{\Sigma}\right)=0. \label{total_rand}
\end{eqnarray}

The users wish to compute a linear function $F$ of the inputs while guaranteeing no information is leaked about another linear function $G$ of the inputs, where $F, G$ are specified as follows. Define ${\bf W} \triangleq [W_1; \cdots; W_K] \in \mathbb{F}_q^{K \times L}$.
\begin{eqnarray}
F = {\bf F} {\bf W} \in \mathbb{F}_q^{M \times L}, ~G = {\bf G} {\bf W} \in \mathbb{F}_q^{N \times L}
\end{eqnarray}
where ${\bf F} \in \mathbb{F}_q^{M \times K}$, ${\bf G} \in \mathbb{F}_q^{N \times K}$ are assumed to have full row rank without any loss, i.e., $M = \mbox{rank}({\bf F}), N = \mbox{rank}({\bf G})$. Further, to avoid degenerate cases, we assume
\begin{eqnarray}
\mbox{matrix ${\bf F}$ does not have zero columns} \label{zero}
\end{eqnarray}
because a zero column, say the $k$-th, means that the input $W_k$ does not appear in $F$, the desired function to compute and then User $k$ can be eliminated from the problem without loss of generality.

To accomplish the goal of vector linear secure aggregation, User $k$ sends to the server a message $X_k$, which is a function of $W_k,Z_k$ and is comprised of $L_X$ symbols from $\mathbb{F}_q$.
\begin{eqnarray}
    H\left(X_k | W_k,Z_k \right) = 0,~\forall k \in [K]. \label{message}
\end{eqnarray}
    
From all messages, the server must be able to recover $F$, with no error.
\begin{eqnarray}
    \mbox{[Correctness]}~~~H\left( F \Bigl| \left( X_k \right)_{k \in [K] }\right) = 0. \label{corr}
\end{eqnarray}
   
Security refers to the constraint that the server cannot infer any additional information about $G$ beyond that contained in $F$.  That is, the following security constraint must be satisfied.
\begin{eqnarray}
    \mbox{[Security]}~~~I\left( G; \left( X_k \right)_{ k \in [K] } \Big| F\right) = 0. \label{security}
\end{eqnarray}
The communication rate $R$, individual key rate $R_Z$ and total key rate $R_{Z_\Sigma}$ characterize how many symbols message $X_k$, key $Z_k$ and source key $Z_\Sigma$ contain per input symbol, and are defined as
\begin{eqnarray}
    R \triangleq \frac{L_X}{L}, ~R_Z \triangleq \frac{L_Z}{L},  ~R_{Z_{\Sigma}} \triangleq \frac{ L_{Z_{\Sigma}}}{L}. \label{rate}
\end{eqnarray}
A rate tuple $(R, R_Z, R_{Z_\Sigma})$ is said to be achievable if there exists a design of keys $Z_k, Z_\Sigma$ and messages $X_k$ of vector linear secure aggregation, for which the correctness constraint (\ref{corr}) and the security constraint (\ref{security}) are satisfied and the communication rate, individual key rate, total key rate are no greater than $R, R_Z, R_{Z_\Sigma}$, respectively. The closure of the set of all achievable rate tuples is called the optimal rate region, denoted as $\mathcal{R}^*$. The main performance metric of this work is the optimal total key rate $R_{Z_\Sigma}^* \triangleq \min_{\mathcal{R}^*} R_{Z_\Sigma}$.

\section{Result}
Our main result is the characterization of the optimal total key rate, stated in the following theorem.
\begin{theorem}\label{thm}
For vector linear secure aggregation defined above, the optimal total key rate is 
\begin{eqnarray}
     R_{Z_{\Sigma}}^* 
    = \mbox{rank} \left( \left[ \mathbf{F} ; \mathbf{G} \right] \right)
     - \mbox{rank} \left( \mathbf{F} \right) = \mbox{rank}({\bf G} | {\bf F}).
\end{eqnarray}
\end{theorem}
The proof of Theorem \ref{thm} is presented in the following two sections.

\section{Achievability Proof}
Let us start with two examples to illustrate the idea in a simpler setting.
\subsection{Example 1: $K = 5, M = 3, {\bf G} = {\bf I}_5$}
Suppose we have $5$ users and wish to compute the following $3$ linear combinations of the inputs over $\mathbb{F}_7$ while security needs to hold for all inputs, i.e., ${\bf G} = {\bf I}_5$, the $5\times 5$ identity matrix.
\begin{eqnarray}
        \mathbf{F}=\begin{bmatrix}
        2 & 0 & 5 & 3 & 1 \\
        5 & 1 & 4 & 2 & 4 \\
        0 & 4 & 3 & 5 & 1  
        \end{bmatrix}
\overset{\mbox{\scriptsize invertible}}{\longrightarrow} 
    \left[\begin{array}{ccc;{2pt/2pt}cc}
    1 & 0 & 0 & 1 & 5\\
    0 & 1 & 0 & 6 & 3\\
    0 & 0 & 1 & 3 & 1
    \end{array} \right] \label{eq:f1}
\end{eqnarray}
where we perform an invertible operation to transform the first $3$ columns of ${\bf F}$ to a $3 \times 3$ identity matrix. Now to achieve the desired total key rate $R_{Z_\Sigma}^* = 5 - 3 = 2$, we set $L = 1$, $L_{Z_\Sigma} = 2$. Suppose we have two uniform and independent noise variables $N_1, N_2$ over $\mathbb{F}_7$, then set
\begin{eqnarray}
    \begin{array}{cc}
        \begin{array}{l}
        X_1  = W_1 - N_1 - 5N_2, \\
        X_2  = W_2 - 6N_1 - 3N_2, \\
        X_3  = W_3 - 3N_1 - N_2,
        \end{array}
    &  \begin{array}{l}
        X_4 = W_4 + N_1, \\
        X_5 = W_5 + N_2 \\
         \\
        \end{array} \\
    \end{array}
\end{eqnarray}
where $X_4, X_5$, corresponding to the last $K-M = 2$ columns of ${\bf F}$, are just the sum of input and each independent noise and $X_1, X_2, X_3$, corresponding to the first $M = 3$ columns of ${\bf F}$, are the sum of input and linear combinations of the noises with coefficients given by the last $2$ columns of ${\bf F}$ after transforming the first $3$ columns to identity (refer to (\ref{eq:f1})).

The scheme is correct because our design allows easy cancellation of the noise to recover $F = {\bf F} {\bf W}$, e.g., to recover the first row of ${\bf F}$, we have $X_1 + X_4 + 5X_5 = W_1 + W_4 + 5W_5$.
To verify security is guaranteed, consider the security constraint (\ref{security}).
\begin{eqnarray}
     I\left( G ; \left( X_k \right)_{k\in[5]} \Big| F \right)
     & = & I\left( \left( W_k \right)_{k\in[5]}; \left( X_k \right)_{k\in[5]} \Big| F \right) \\
     & = & H\left( \left( X_k \right)_{k\in[5]} \Big| F \right) -H \left( \left( X_k \right)_{k\in[5]} \Big| \left( W_k \right)_{k\in[5]}, F \right) \\
     & = &  H\left( \left( X_k \right)_{k\in[5]} , F \right) - H(F) - H\left(N_1, N_2 \Big| \left( W_k \right)_{k\in[5]} \right) \\
     & \leq & 5 - 3 - 2 = 0
\end{eqnarray}
where the last step follows from the fact that messages $(X_k)_k$ contain $5$ symbols, $F$ can be recovered from $(X_k)_k$, and uniform key symbols $N_1, N_2$ are independent of the inputs $W_k$ (refer to (\ref{h1})). 

\subsection{Example 2: $K=6, M = 2, N = 3$}
Suppose we have $6$ users and wish to compute the following $2$ linear combinations of the inputs over $\mathbb{F}_7$ while revealing nothing about the following $3$ linear combinations of the inputs.
\begin{align}
    \begin{array}{cc}
        \mathbf{F} = 
        \begin{bmatrix}
        1 & 0 & 5 & 5 & 3 & 5\\
        0 & 1 & 5 & 6 & 0 & 3
        \end{bmatrix},
    &  
        \mathbf{G} = \begin{bmatrix}
            3 & 0 & 1 & 4 & 2 & 4\\
            2 & 2 & 1 & 3 & 5 & 3\\
            1 & 1 & 3 & 4 & 3 & 1
        \end{bmatrix} \\
    \end{array}
\end{align}
where we have assumed without loss of generality that ${\bf F}$ contains an identity matrix as a submatrix (as ${\bf F}$ has full rank). It turns out that $\mbox{rank}({\bf G} | {\bf F}) = 2$; in particular, the last row of ${\bf G}$ is equal to the sum of the two rows of ${\bf F}$ while the first two rows of ${\bf G}$ are linearly independent of ${\bf F}$.

Following the idea of Example 1, we may set $L = 1$ and
\begin{eqnarray}
    \begin{array}{ll}
        X_1 = W_1 - 5N_1 - 5N_2 - 3N_3 - 5N_4, & X_2 = W_2  -5N_1 - 6N_2 -3N_4,\\
        X_3 = W_3 + N_1, & X_4 = W_4 + N_2, \\
        X_5 = W_5 + N_3, & X_6 = W_6 + N_4,
    \end{array} \label{eq:x1}
\end{eqnarray}
but we use more key symbols $(N_1, N_2, N_3, N_4)$ than the optimum ($2 = \mbox{rank}({\bf G} | {\bf F})$). It turns out that $N_1, N_2, N_3, N_4$ need not to be independent and the following step of introducing correlation among them is the most technical step of the proof.

Find a $2 \times 4$ matrix ${\bf V}_{2\times 4}$ such that 
\begin{eqnarray}
        \left[ \begin{array}{c}
        {\mathbf F} \\
        {\mathbf G} \\        
        \hdashline[4.8pt/4.8pt]
        \begin{array}{c: c}
        \begin{matrix}
          0 & 0  \\
         0 & 0  
        \end{matrix} & 
         \ {\mathbf{V}} 
         \end{array}\\ 
        \end{array}
        \right]
        ~\mbox{has full rank $6$.} \label{eq:full}
\end{eqnarray}

Note that such ${\bf V}$ exists as $\mbox{rank}([{\bf F}; {\bf G}]) = 4$. Any such ${\bf V}$ will work and then we find its right null space ${\bf V}^\perp_{4\times 2}$ such that ${\bf V}{\bf V}^{\perp} = {\bf 0}_{2 \times 2}$ (the $2\times 2$ all zero matrix, again any one will work). For example, we may set
\begin{eqnarray}
    \mathbf{V}=\begin{bmatrix}
        1 & 0 & 6 & 6 \\
        0 & 1 & 6 & 5
    \end{bmatrix}, ~
        {\bf V}^\perp =\begin{bmatrix}
        1 & 1 \\
        1 & 2 \\
        1 & 0 \\
        0 & 1
    \end{bmatrix}.
\end{eqnarray}
Then the noise symbols $N_1, N_2, N_3, N_4$ can be generated from two uniform and independent noise symbols $S_1, S_2$ by precoding with ${\bf V}^{\perp}$,
\begin{eqnarray}
    \begin{bmatrix}
        N_1 \\
        N_2 \\
        N_3 \\
        N_4
    \end{bmatrix} =
    {\bf V}^\perp \times \begin{bmatrix}
        S_1 \\
        S_2
    \end{bmatrix} = \begin{bmatrix}
        S_1 + S_2 \\
        S_1 + 2S_2\\
        S_1 \\
        S_2
    \end{bmatrix}. \label{eq:s1}
\end{eqnarray}
We may write out the final message assignment with noise symbols $S_1, S_2$.
\begin{eqnarray}
    \begin{array}{ll}
        X_1 = W_1 - 6S_1 -6S_2, & X_2 = W_2 - 4S_1 - 6S_2, \\
        X_3 = W_3 + S_1 + S_2,  &  X_4 = W_4 + S_1 + 2S_2, \\
        X_5 = W_5 + S_1, &       X_6 = W_6 + S_2.
    \end{array} \label{eq:x2}
\end{eqnarray}

The scheme is correct due to the same reasoning as that of Example 1 (refer to (\ref{eq:x1}) for noise cancellation). Let us verify that the security constraint (\ref{security}) is satisfied.
\begin{eqnarray}
    && I\left( G; \left( X_k \right)_{ k \in[6]} \Bigl| F \right) \notag \\
    & =  & H \left( \left( X_k \right)_{k\in[6]} \Bigl| F \right)-H \left( \left( X_k \right)_{k\in[6]} \Bigl| G, F \right) \\
    & =  & \left[H\left( \left(X_k \right)_{k\in[6]}, F \right) - H(F)\right] - H\left(\left( X_k \right)_{k\in[6]}, {\bf V}[X_3; X_4; X_5; X_6] \Bigl| G, F \right) \\
    & \leq  & (6 - 2) - H\left(\left( X_k \right)_{k\in[6]}, {\bf V}[W_3; W_4; W_5; W_6] \Bigl| G, F \right) \label{eq:t1} \\
    & = & 4 - H\left( {\bf V}[W_3; W_4; W_5; W_6] \Bigl| G, F \right) - H\left(\left( X_k \right)_{k\in[6]} \Bigl| {\bf V}[W_3; W_4; W_5; W_6], G, F \right) \\
    &=& 4 - 2 - H(S_1, S_2) \label{eq:t2} \\
    & = & 4 - 2 -2 = 0
\end{eqnarray}
where (\ref{eq:t1}) follows from ${\bf V}{\bf V}^\perp = {\bf 0}$, i.e., the noise variables in $X_3, X_4, X_5, X_6$ (precoded by ${\bf V}^\perp$, refer to (\ref{eq:s1}), (\ref{eq:x2})) are zero forced after multiplying ${\bf V}$. In (\ref{eq:t2}), we use the fact that ${\bf V}[W_3; W_4; W_5; W_6]$, ${\bf G} {\bf W}$, ${\bf F} {\bf W}$ have full rank and are thus invertible to $W_1, \cdots, W_6$ (refer to (\ref{eq:full})).

\subsection{General Achievability Proof} \label{sec:ach}
The achievability proof for the general case is an immediate generalization of the above examples. 

Without loss of generality, assume 
\begin{eqnarray}
{\bf F} =    \left[\begin{array}{c;{2pt/2pt}c}
    \mathbf{I}_M  & \tilde{\mathbf{F}}_{M \times (K-M)} 
    \end{array} \right] \label{eq:ff}
\end{eqnarray}
as otherwise we may relabel the columns (user indices) and perform an invertible operation to obtain the above form.

Find\footnote{We may find any $K - \mbox{rank}([{\bf F}; {\bf G}])$ row vectors that are linearly independent of $[{\bf F}; {\bf G}]$ and then use the first $M$ columns of ${\bf F}$ (i.e., the identity submatrix) to eliminate the first $M$ columns of these row vectors to obtain {\bf V}.} a $(K - \mbox{rank}([{\bf F}; {\bf G}])) \times (K-M)$ matrix ${\bf V}$ such that 
\begin{eqnarray}
    \left[\begin{array}{c}
        \mathbf{F} \\ 
        \mathbf{G} \\\hdashline[2pt/2pt]
        \begin{array}{c;{2pt/2pt}c}
            \mathbf{0}_{ (K - \mbox{\scriptsize rank}([{\bf F}; {\bf G}])) \times M } & \mathbf{V} 
        \end{array}
    \end{array}\right]
        ~\mbox{has full rank $K$} \label{eq:full1}
\end{eqnarray}
where $\mathbf{0}_{a \times b}$ is the $a \times b$ all zero matrix. As $\mbox{rank}({\bf V}) = K - \mbox{rank}([{\bf F}; {\bf G}])$, we may assume without loss of generality that (as otherwise may obtain through an invertible operation)
\begin{eqnarray}
{\bf V} = 
\left[\begin{array}{c;{2pt/2pt}c}
    \mathbf{I}_{K - \mbox{\scriptsize rank}([{\bf F}; {\bf G}])}  & \tilde{\mathbf{V}}_{ (K - \mbox{\scriptsize rank}([{\bf F}; {\bf G}])) \times (\mbox{\scriptsize rank}([{\bf F}; {\bf G}])-M)}
    \end{array} \right]
\end{eqnarray}
and set\footnote{When $K = \mbox{rank}([{\bf F}; {\bf G}])$, ${\bf V}$ is null and we set ${\bf V}^\perp = {\bf I}_{K-M}$.}
\begin{eqnarray}
    {\bf V}^\perp = 
    \left[\begin{array}{c}
        -\tilde{\mathbf{V}} \\\hdashline[2pt/2pt]
        \mathbf{I}_{\mbox{\scriptsize rank}([{\bf F}; {\bf G}])-M}
    \end{array}\right]_{(K-M) \times (\mbox{\scriptsize rank}([{\bf F}; {\bf G}])-M)} \label{eq:vperp}
\end{eqnarray}
so that
\begin{eqnarray}
    {\bf V} {\bf V}^\perp = -\tilde{\bf V} + \tilde{\bf V} = {\bf 0}.
\end{eqnarray}

We are now ready to describe the secure aggregation protocol. Set $L = 1, L_{Z_\Sigma} = \mbox{rank}([{\bf F}; {\bf G}]) - \mbox{rank}({\bf F}) = \mbox{rank}([{\bf F}; {\bf G}]) - M$ (thus the total key rate $R_{Z_\Sigma}$ achieves the optimum in Theorem \ref{thm}). Consider $L_{Z_\Sigma}$ uniform and independent noise symbols ${\bf S} \triangleq [S_1; \cdots; S_{L_{Z_\Sigma}}]$ and set
\begin{eqnarray}
    {\bf N} \triangleq [N_1; \cdots; N_{K-M}] &=& {\bf V}^{\perp} {\bf S}, \notag \\
    \left[ X_1; \cdots; X_M \right] &=& \left[ W_1; \cdots; W_M \right] - \tilde{\bf F} [N_1; \cdots; N_{K-M}], \label{eq:scheme}\\
    \left[ X_{M+1}; \cdots; X_K \right] &=& \left[ W_{M+1}; \cdots; W_K \right] + [N_1; \cdots; N_{K-M}]. \notag
\end{eqnarray}

Let us prove the above scheme is correct and secure. For correctness (refer to (\ref{corr})), we have 
\begin{eqnarray}
    F = {\bf F} {\bf W} &\overset{(\ref{eq:ff})}{=}& \left[ W_1; \cdots; W_M \right] + \tilde{\bf F} \left[ W_{M+1}; \cdots; W_K \right] \\
    &\overset{(\ref{eq:scheme})}{=}& \left[ X_1; \cdots; X_M \right] + \tilde{\bf F} \left[ X_{M+1}; \cdots; X_K \right]
\end{eqnarray}
so that $F$ can be decoded correctly from $(X_k)_k$. For security (refer to (\ref{security})), we have
\begin{eqnarray}
     && I\left( G; \left( X_k \right)_{k\in[K]} \Bigl| F \right) \notag \\
    &=& H \left( \left( X_k \right)_{k\in[K]}\Bigl| F \right)-H \left( \left( X_k \right)_{k\in[K]}\Bigl| G,F \right) \\
    &=& H \left( \left( X_k \right)_{k\in[K]}, F \right) - H(F) -H \left( \left( X_k \right)_{k\in[K]}, {\bf V} \left[X_{M+1}; \cdots; X_K \right] \Bigl| G,F \right) \\
    &\overset{(\ref{corr})(\ref{eq:scheme})}{=}& H \left( \left( X_k \right)_{k\in[K]} \right) - H(F) -H \left( \left( X_k \right)_{k\in[K]}, {\bf V} \left[W_{M+1}; \cdots; W_K\right] \Bigl| G,F \right) \label{eq:tt} \\
    &\leq& (K - M) - H \left( {\bf V} [W_{M+1}; \cdots; W_K] \Bigl| G,F \right) \notag\\
    &&~ - H \left( \left( X_k \right)_{k\in[K]} \Bigl| {\bf V} [W_{M+1}; \cdots; W_K], G,F \right) \\
    &\overset{(\ref{eq:full1})}{=}& (K-M) -  \left[ K-\mbox{rank}(\mathbf{[F;G]}) \right] - H \left( \left( X_{k} \right)_{k\in[K]}\Bigl| \left( W_k \right)_{k\in[K]} \right)  \\
    &\overset{(\ref{eq:scheme})}{=}& \mbox{rank}(\mathbf{[F;G]}) - M - H \left( {\bf N} \Bigl| \left( W_k \right)_{k\in[K]} \right)  \\
    &\overset{(\ref{eq:scheme})}{=}& \mbox{rank}(\mathbf{[F;G]}) - M - H \left( {\bf V}^\perp {\bf S} \Bigl| \left( W_k \right)_{k\in[K]} \right)  \\
    &\overset{(\ref{eq:vperp})}{=}& \mbox{rank}(\mathbf{[F;G]}) - M - H \left( {\bf S} \Bigl| \left( W_k \right)_{k\in[K]} \right)  \\
    &=& \mbox{rank}(\mathbf{[F;G]}) - M - L_{Z_\Sigma}  \\
    &=& \mbox{rank}(\mathbf{[F;G]}) - M - \left[\mbox{rank}(\mathbf{[F;G]}) - M\right] = 0.
\end{eqnarray}

\begin{remark}
Our code design is guided by the above security proof, where the key is to ensure that $H \left( \left( X_k \right)_{k\in[K]}\Bigl| G,F \right)$ is maximized and is equal to $K-M$ (equivalent to that nothing is revealed about $G$). To this end, we wish to ensure that in $(X_k)_k$, we maximize the size of the part that is just functions of inputs $W_k$ and is independent of $G,F$ (extracted through ${\bf V}$ in (\ref{eq:tt})) so that the noise needs to be beamformed along ${\bf V}^\perp$, i.e., orthogonal to ${\bf V}$ and in this way, the amount of noise used is minimized. This is the rationale behind the design of noise (particularly ${\bf V}, {\bf V}^\perp$) and what we mean that our scheme is reverse engineering the mutual information term being zero in the security constraint (\ref{security}).
\end{remark}

\section{Converse Proof}\label{Conv}
Let us start with two useful lemmas.
First, we show that the messages cannot reveal too much information about the inputs as otherwise the security constraint (\ref{security}) will be violated.

\begin{lemma}\label{lemma:leak}
The following inequality holds.
    \begin{eqnarray}
        I\left( \left(X_k \right)_{k\in[K]}; \left( W_k \right)_{k\in[K]}\ \right) \leq \left[K - \mbox{rank}({\bf G} | {\bf F})\right] L.
    \end{eqnarray}
\end{lemma}

{\it Proof:}
\begin{eqnarray}
    && \ I\left( \left( X_k \right)_{ k\in[K]} ; \left( W_k \right)_{k\in[K]} \right) \notag \\
    &= & \ I\left( \left( X_k \right)_{k\in[K] }; \left( W_k \right)_{ k\in[K]}, G \right) \\
     &= & \ I\left((X_k)_{k\in[K]}; G \right) + I\left( \left( X_k \right)_{ k\in[K] }; \left( W_k \right)_{k\in[K]} \Bigl| G \right) \\
      &\leq& \ I \left( \left( X_k \right)_{k\in[K]}, F; G \right) + H \left( \left( W_k \right)_{k\in[K]} \Bigl| G \right) \label{eq:c1}\\
       &= & \ I( F; G) + \underbrace{I\left(  G; \left( X_k \right)_{k\in[K]} \Bigl| F \right)}_{\overset{(\ref{security})}{=}0} +~ H \left( \left( W_k \right)_{k\in[K]}, G \right) - H(G)  \\
       &\leq & \ \left[\mbox{rank}({\bf G}) - \mbox{rank}({\bf G} | {\bf F}) \right] L + [K - \mbox{rank}({\bf G})]L \\
       &= & \ 
       \left[K - \mbox{rank}({\bf G} | {\bf F}) \right]L.
\end{eqnarray}

\hfill\QED

Second, we show that each user needs to send its input fully to the server as $F$ contains each input. Define $A\backslash B$ as the set of elements that belong to set $A$ but not set $B$ and ${\bf A}(:, u)$ as the $u$-th column of matrix ${\bf A}$.
\begin{lemma}\label{lemma:xk}
For any $u \in [K]$, we have
\begin{eqnarray}
H\left(X_u|(W_k,Z_k)_{k\in[K]\backslash \{u\}}\right) \geq L. 
\end{eqnarray}
\end{lemma}

{\it Proof:} Consider any $u \in [K]$.
\begin{eqnarray}
    && H\left(X_u|(W_k,Z_k)_{k\in[K]\backslash \{u\}}\right)\notag\\
    &\geq& I\left(X_u; F \Big|(W_k,Z_k)_{k\in[K]\backslash \{u\}}\right)\\
    &=& H\left( F \Big| (W_k,Z_k)_{k\in[K]\backslash \{u\}}\right) - H\left( F \Big| X_u, (W_k,Z_k)_{k\in[K]\backslash \{u\}}\right)\\
    &\geq& H\left({\bf F}(:,u) W_u\right) - 
    H\left( F \Big| (X_k)_{k\in[K]}\right) \label{eq:c2} \\
    &=& H(W_u) - 0 \label{eq:c3} \\
    &=& L
\end{eqnarray}
where in (\ref{eq:c2}), the first term follows from the independence of the inputs $(W_k)_k$ and the keys $(Z_k)_k$ (refer to (\ref{h1})) and the second term is due to the fact that message $X_k$ is a function of input $W_k$ and key $Z_k$ (refer to (\ref{message})), and dropping conditioning cannot decrease entropy. In (\ref{eq:c3}), the first term follows from the fact that ${\bf F}$ does not contain zero columns so that ${\bf F}(:, u)$ is not the zero vector (refer to (\ref{zero})) and the second term is due to the correctness constraint (\ref{corr}).

\hfill\QED

\begin{remark}
The optimal communication rate of vector linear secure aggregation is $\min_{\mathcal{R}^*} R = 1$ as the converse follows from Lemma \ref{lemma:xk} and the achievability follows from the scheme in Section \ref{sec:ach}.
\end{remark}

We are now ready to proceed to the converse proof.  
\begin{eqnarray}
    L_{Z_{\Sigma}} &\ \geq \ & \ H(Z_{\Sigma}) \label{eq:cc} \\
    &\geq \ & \ H\left(\left( Z_k \right)_{k\in[K]}\right) \\
    & \geq \ & \ I\left(\left( Z_k \right)_{k\in[K]};\left(X_k\right)_{k\in[K]} \Bigl|\left(W_k\right)_{k\in[K]}\right) \\
    & = \ & \ I\left( \left( Z_k \right)_{k\in[K]};(X_k)_{k\in[K]} \Bigl|(W_k)_{k\in[K]}\right) + H\left(\left( X_k \right)_{k\in[K]} \Bigl|\left( W_k,Z_k \right)_{k\in[K]}\right) \\
    & = \ & \ H\left(\left(X_k\right)_{k\in[K]} \Bigl|(W_k)_{k\in[K]}\right) \\
    & = \ & \ H\left(\left( X_k \right)_{k\in[K]}\right) - I\left(\left(X_k\right)_{k\in[K]};(W_k)_{k\in[K]}\right) \\
    & \geq \ & \sum_{u=1}^K H\left(X_u \Bigl| (W_k,Z_k)_{k\in[K]\backslash \{u\}}\right) - I\left(\left(X_k\right)_{k\in[K]};(W_k)_{k\in[K]}\right)\\
    &\geq& KL - \left[K - \mbox{rank}({\bf G} | {\bf F})\right] L = \mbox{rank}({\bf G} | {\bf F}) L \label{eq:c4}\\ 
    &\Rightarrow& R_{Z_{\Sigma}}= \  \ \frac{L_{Z_{\Sigma}}}{L} ~\geq~ \mbox{rank}({\bf G} | {\bf F})
\end{eqnarray}
where (\ref{eq:c4}) follows from Lemma \ref{lemma:leak} and Lemma \ref{lemma:xk}.

\section{Discussion} \label{sec:dis}
While the optimal communication rate and total key rate of vector linear secure aggregation is fully solved, the optimal individual key rate remains elusive. Here we give a simple open setting. Consider $K = 3$ users, the function to compute is $F = W_1 + W_2 + W_3$, and the function to protect is $G = W_1 + 2W_2 + 3W_3$, defined over $\mathbb{F}_5$ (the exact field is not important, here just to ensure that $1,2,3$ are defined over the field). The best existing result on the individual key rate $R_Z$ is that $1/2 \leq R_Z \leq 2/3$ but it is unknown if the upper or lower bound is tight. The converse result of $R_Z \geq 1/2$ follows\footnote{Following (\ref{eq:cc}), we may show that $2L_Z \geq H(Z_1, Z_2) \geq L$, where in Lemma \ref{lemma:leak}, the inequality to prove can be replaced by $I(X_1, X_2; W_1, W_2) \leq L$ and $G$ can be replaced by $\tilde{G} = 3F-G = 2W_1 + W_2$, i.e., the part of $W_1, W_2$ that needs to be protected is $\tilde{G}$ so that $X_1, X_2$ cannot reveal more than $L$ bits of information about $W_1, W_2$.} with minor modifications from the total key rate converse proof and the achievability result of $R_Z \leq 2/3$ follows\footnote{The proof of correctness and security is similar to that of Theorem \ref{thm}. By permuting, we will replicate (\ref{eq:open}) $L = 3$ times and each time one user just sends out its input without being covered by any noise so that each user employs $2$ noise symbols, i.e., $L_Z = 2$ and $R_Z = L_Z/L = 2/3$ is achieved.} by permuting and symmetrizing the following simple scalar scheme ($L=1$, $N$ is a uniform random noise symbol)
\begin{eqnarray}
X_1 = W_1, X_2 = W_2 + N, X_3 = W_3 - N. \label{eq:open}
\end{eqnarray}

The generalization from considering sum computation to more general functions (specifically linear) has appeared in recent related information theoretic context such as coded caching \cite{wan2020linear}, private information retrieval \cite{gholami2024multi, zhu2024private}, and distributed computation \cite{wan2020distributed, malak2024multi, Sun_Jafar_CB, yao2024generic}. The generalization from perfect security to weaker secrecy metrics has appeared in secure summation literature, where leakage about all inputs is allowed but amount bounded by an explicit parameter \cite{chou2024private}, leakage is prevented about some subset of inputs (with multiple such security sets and multiple sets of colluding servers) \cite{li2023weakly}, and in network coding for computing \cite{Xuan_Yang_Yeung_Source, Xuan_Yang_Yeung_Function, xu2022network}. Finally, an interesting research avenue is to explore the synergistic benefits of other elements such as uncoded keys \cite{Zhao_Sun_Sum, wan2024uncoded} and more than one hop network topologies \cite{sun2023secure, zhang2024optimal}.

\bibliography{Thesis}

\end{document}